\documentclass[floatfix,aps,prl,twocolumn,showpacs,preprintnumbers,amsmath,amssymb]{revtex4}
\usepackage{epsfig}
\usepackage{graphicx}
\usepackage{dcolumn}
\usepackage{bm}
\usepackage{ulem}
\usepackage{color}

\begin{document}
\title{Effect of magnetic Gd impurities on superconductivity in MoGe films
with different thickness
and morphology}
\author{Hyunjeong Kim, Anil Ghimire, Shirin Jamali, Thaddee K. Djidjou,
Jordan M. Gerton, and A. Rogachev}
\affiliation{Department of Physics and Astronomy, University of
Utah, Salt Lake City, Utah 84112, USA}
\date{\today}

\begin {abstract}
We studied the effect of magnetic doping with Gd atoms on the
superconducting properties of amorphous Mo$_{70}$Ge$_{30}$ films. We
observed that in uniform films deposited on amorphous Ge, the
pair-breaking strength per impurity strongly decreases with film
thickness initially and saturates at a finite value in films with
thickness below the spin-orbit scattering length. The variation is
likely caused by surface induced magnetic anisotropy and is
consistent with the fermionic mechanism of superconductivity
suppression. In thin films deposited on SiN the pair-breaking
strength becomes zero. Possible reasons for this anomalous response
are discussed. The morphological distinctions between the films of
the two types were identified using atomic force microscopy with a
carbon nanotube tip.

\end {abstract}


\maketitle
 Understanding
physical processes related to localized magnetic moments is
particularly important for low-dimensional systems since such
moments can form spontaneously on surfaces and interfaces of
nominally non-magnetic materials. The formation of localized
magnetic moments is well known in semiconductor heterostructures and
devices, where they are carried by structural defects with unpaired
electrons \cite{Poindexter}. Localized magnetic moments were
recently detected on the surface of a normal metal \cite{Bluhm}; in
superconducting systems, they are believed to be responsible for
several unusual effects such as 1/f noise in SQUIDs and qubits
\cite{McDermott} and an anomalous magnetic field enhancement of a
critical current in nanowires \cite{Rogachev}. The origin of
spontaneously formed magnetic moments often remains unknown; on the
other hand, their effects can be probed by magnetic moments that are
introduced intentionally.

Here we study the effect of intentional magnetic doping on transport
properties of ultra-thin MoGe films that undergo a
superconductor--insulator transition (SIT)
\cite{Heviland,Markovic1,Gantmakher,Shahar,Baturina1, Kapitulnik1}.
The mechanism of the SIT remains an important unresolved problem in
condensed matter physics. In general, there are several distinct
physical processes that may lead to the SIT. Within the fermionic
mechanism, Cooper pairing is locally suppressed by disorder-enhanced
electron-electron repulsion \cite{Finkelstein1,Kirkpartick}. The
fermionic theories predict that the pair-breaking strength of
magnetic impurities does not change with increasing disorder or
decreasing film thickness \cite{Belitz,Smith}. Experimentally,
magnetic doping was studied in quench-condensed Pb\cite{Parker1} and
Pb-Bi films \cite{Chervenak1}. In the latter case, behavior
consistent with the fermionic theory was observed relatively far
from the SIT. Several bosonic mechanisms were proposed for the
critical regime of the SIT. In these models, Cooper pairs are
preserved across the transition but coherence in the films is lost
due to vortex proliferation \cite{Fisher1}, disorder-induced Cooper
pair localization \cite{Feigelman1,Pokrovsky1}, or fluctuations of
the superfluid order parameter \cite{Spivak1}. While the models
cited in Ref. \cite{Feigelman1,Pokrovsky1,Spivak1} differ in their
detail microscopic mechanisms, they all predict the appearance of a
spatially inhomogeneous superconducting state. The emergence of this
state was observed in numerical simulations \cite{Ghosal} and
 was recently detected experimentally \cite{Sacepe1}.
Possible effects of magnetic pair-breaking within the bosonic models
have not yet been analyzed theoretically.

The amorphous MoGe system is particular suitable for studying
magnetic doping. This is the only known system where suppression of
the critical temperature can be explained by the fermionic theory in
all range of films thicknesses. Moreover, this can be achieved with
the constrained theory, which assumes that effective electron-phonon
coupling is not affected by disorder or film thickness. MoGe films
with this property need to be deposited on a substrate covered with
an underlayer of amorphous Ge that helps to maintain constant bulk
resistivity of the film and ensures its homogeneity \cite{Graybeal}.
On the other hand, a missing Ge underlayer makes it possible to
obtain and test an inhomogeneous superconducting state. We selected
Gd as a magnetic dopant because its magnetic moment is carried by a
half-filled $f$-shell and does not depend on the host material.

The critical temperature of amorphous Mo$_x$Ge$_{100-x}$ alloys
depends on the particular value of $x$. In the first stage of our
study, we used co-sputtering from three independently controlled
guns with Mo, Ge, and Gd targets to fabricate a series of thick
Mo$_x$Ge$_{100-x}$-Gd films with varying Gd content and $x$ in the
range 50-80 at. $\%$. From transport measurements on these films we
found that the alloy with $x\simeq70$ is the most suitable for the
Gd doping. In this alloy the superconductivity is completely
suppressed when 6.5 at. $\%$ of Gd is added; at lower Gd content we
detected a single-step superconducting transition in $R(T)$ curves.

We fabricated two series of Mo$_{70}$Ge$_{30}$ films. Films of the
A-series were deposited on a Si substrate covered with a 60-nm thick
layer of SiN grown by chemical vacuum deposition. For the B-series,
prior to the deposition of MoGe film, a 3-nm thick underlayer of
amorphous Ge was deposited. For oxidation protection the films of
both series were covered by a 3-nm thick layer of Ge.

\begin{figure}
\begin{center}
\includegraphics[width = 3.0in]{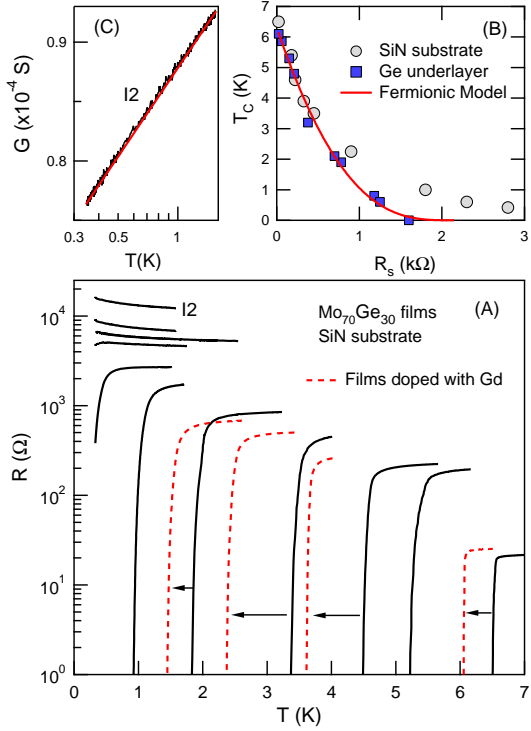}
\caption{\label{fig:SIT} (A) Sheet resistance versus temperature for
a series of amorphous Mo$_{70}$Ge$_{30}$ films deposited on SiN
substrates (solid lines). The dashed lines indicate films doped with
Gd; arrows indicate the correspondence between undoped and doped
films.(B) Critical temperature versus sheet resistance for two
series of undoped Mo$_{70}$Ge$_{30}$ films. The red solid line
indicates a fit to the fermionic model. (C) Conductance of the
insulating film I2 as a function of temperature on a logarithmic
scale. The solid red line is a linear fit.}
\end{center}
\end{figure}

In Fig. 1A, we show the temperature dependence of the sheet
resistance for undoped Mo$_{70}$Ge$_{30}$ films deposited on SiN
(A-series). Within the studied temperature range (down to 0.3 K) the
system undergoes a direct SIT with no intermediate metallic phase.
As shown in Fig. 1C, in the insulating regime, conductance has the
logarithmic temperature correction arising due to the weak
localization and electron-electron interaction contributions.
Qualitatively similar suppression of superconductivity was observed
for films deposited on Ge underlayer. Figure 1B shows the mean-field
$T_c$ (defined at the middle of the transition) as a function of
sheet resistance for the two series of films. As expected we found
that the $T_c$ for the B-series can be well fitted by the fermionic
model \cite{Finkelstein1}. However the $T_c$ of the A-series
deviates from the model for the films with thickness below 1.5 nm.
It is interesting to note that the critical sheet resistance of the
A-series ($\approx$ 5 k$\Omega$) is close to the universal sheet
resistance $R_q=h/4e^2=6.45$ k$\Omega$ predicted within the ``dirty
boson'' model \cite{Fisher1} The difference between A- and B-series
cannot be explained by change in the dielectric constant of the
substrate. SiN has lower dielectric constant than $\alpha$-Ge;
therefore, electron-electron interactions in this system  have worse
screening and suppression of $T_c$ would be expected at lower values
of sheet resistance than in B-series.

\begin{figure}
\begin{center}
\includegraphics[width = 3.0in]{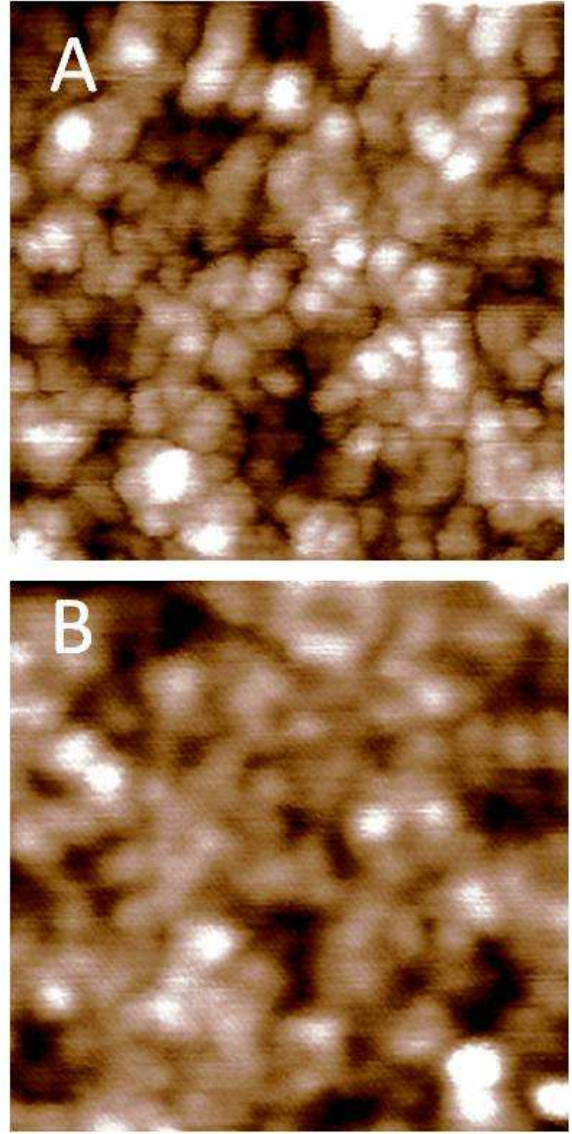}
\caption{\label{fig:AFM} (A) 2D atomic force microscopy image of the
surface of 1-nm thick MoGe film deposited on SiN (A-type). (B) 2D
AFM image of a similar film deposited on SiN covered with 3-nm thick
Ge underlayer (B-type). Size of the images is $200\times200$ nm$^2$}
\end{center}
\end{figure}

\begin{figure}
\begin{center}
\includegraphics[width = 3.4in]{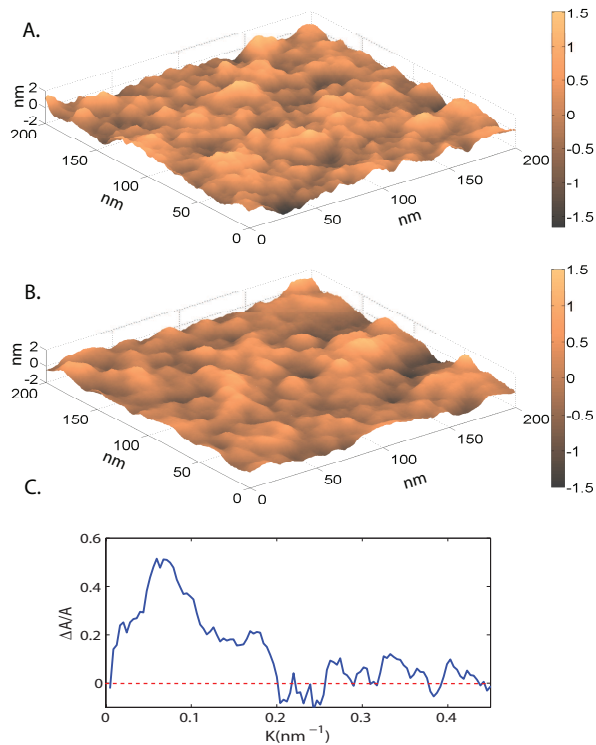}
\caption{\label{fig:AFM} (A) 3D atomic force microscopy image of the
surface of 1-nm thick MoGe film deposited on SiN (A-type). (B) 3D
AFM image of a similar film deposited on SiN covered with 3-nm thick
Ge underlayer (B-type). (C) The difference $\Delta A$ between the 1D
Fourier profiles of the AFM images, normalized by the average
profile, $A$, for the A-type and B-type films.}
\end{center}
\end{figure}

Looking for a possible structural effect of the Ge underlayer, we
inspected surface morphology of several test samples with an atomic
force microscope (AFM) equipped with a carbon nanotube tip that
provided 2 nm lateral resolution. Typical 2D images are shown in
Fig. 2A,B and 3D AFM images in Fig3A,B. To extract quantitative
information about the lateral scale of the topographical features, a
spatial Fourier transform was performed on the AFM  data.  To reduce
noise, the resulting data were averaged over different directions in
Fourier space, yielding a 1D profile of $A$ vs. $k$ for both the SiN
and Ge underlayer samples. In Fig. 3C we show the difference between
the 1D Fourier profiles of the A-type and B-type films shown in Fig.
3A and Fig. 3B, normalized to the average of the A-type and B-type
profiles.

The overabundance of intermediate spatial frequencies for the type-A
sample indicates that the Ge underlayer smoothes the surface and
suppresses topographical features with a characteristic lateral
scale of about 15 nm. This smoothing of the surface is also evident
in comparing the AFM images. On the other hand, the \textit{average}
surface roughness, which characterizes the \textit{height} of the
topographical features, is 0.4-0.5 nm for both A-type and B-type
films. Evidently, the missing Ge underlayer introduces
inhomogeneities that are not strong enough to form a granular
structure. This is also evident from transport measurements. We see
from Fig. 1A that even for thinnest films the superconducting
transition remains sharp with a well-defined $T_c$. There is no tail
in $R(T)$ as it is typically observed in granular materials
\cite{Herzog}. Normal state properties of the films also do not
indicate the presence of strong inhomogeneities. From the theory of
weak localization (eq. 4.47a in Ref. \cite{Altshuler}) we estimated
that the dephasing length $L_\varphi$ at $T$=0.3 K is about 80 nm
for our least resistive insulating film. Since we do not see any
sign of the insulating behavior down to $T$=0.3 K, the one-electron
localization length in our films should be larger (probably much
larger) than 80 nm and thus cover many random ``hills'' and
``valleys'' of the films' morphological profile.

The inset to Fig. 4A shows an arrangement of samples and targets
used for fabrication of the MoGe films doped with Gd. The deposition
was carried out by co-sputtering from two guns. Several substrates
were positioned approximately at the same distance from the
composite MoGe target, but at varying distances from the Gd target.
For each position, the deposition rate was calibrated by profile
measurement of a test thick film; the thickness of films was
controlled by the deposition time. Films within each series were
fabricated in the same run under the same vacuum and deposition
conditions. They have the same thickness but systematically varying
Gd content.

\begin{figure}[b]
\begin{center}
\includegraphics[width = 3.0in]{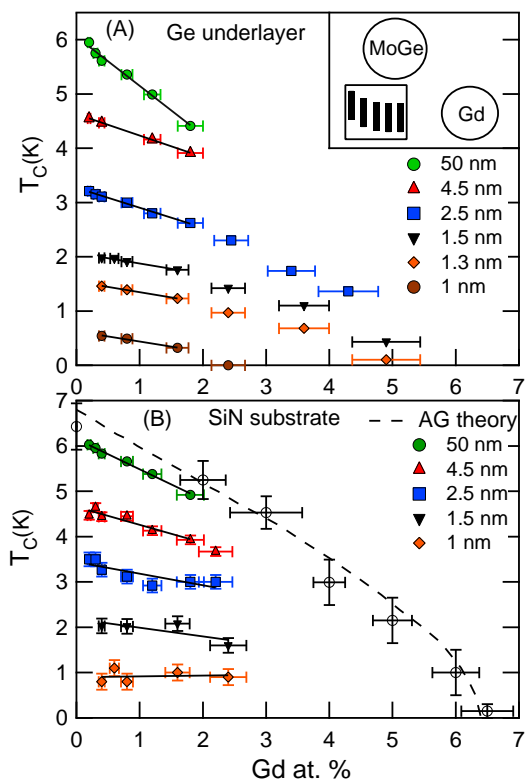}
\caption{\label{fig:TcGd} (A) Critical temperature versus Gd content
for films with the Ge underlayer. The inset indicates the
arrangements of samples and targets in the sputtering chamber for
two-gun co-sputtering (B) Critical temperature versus Gd content for
films with indicated thickness deposited on the SiN substrates. The
dashed line is a fit of the data in the extended Gd percentage range
to the Abrikosov-Gor'kov theory.   }
\end{center}
\end{figure}

Temperature dependence of sheet resistance for several
representative Gd-doped films is shown as dashed lines in Fig. 1A.
The magnetic doping simply shifts the superconducting transition,
leaving its width and normal state resistance essentially unchanged.
Figure 4 displays the critical temperature versus Gd content. Errors
in the Gd content originate from uncertainty in the time of the
deposition, deposition rate and positioning of a sample holder
inside of the chamber. In addition, we found that thin A-type films
with the same nominal thickness deposited in the same run revealed
random variations in $T_c$. It is interesting to note that this
effect was not detected in the B-type films deposited on Ge. In
A-type films random deviations of $T_c$ from the average value were
always accompanied by the change in the normal state $R_s$ of a
film; in fact, there was no uncertainty in $T_c$ vs $R_s$ relation.
The error in $T_c$ resulting from this effect is indicated by
vertical error bars in Fig. 4B.  It was estimated from measurements
on several series of undoped films deposited in the same conditions
as the doped ones.

In Fig. 4B we showed with open circulus the dependence of the $T_c$
on Gd content for the thick MoGe films fabricated by three-gun
(Mo,Ge and Gd) deposition in the extended range of doping . The
dependence can be fitted by the Abrikosov-Gor'kov (AG) theory
\cite{AG}. The critical concentration of Gd is 6.5 at. $\%$; the
corresponding volume critical concentration is $n_c=3\times10^{21}$
cm$^{-3}$. The rest of the data were obtained with the two-gun
deposition. The AG theory predicts that at low doping, $T_c$ behaves
as $k_B(T_{c0}-T_{c})=\pi \alpha/4$. The total pair-breaking
strength, $\alpha$, is related to the pair-breaking strength per
impurity as $\alpha=\alpha_p n_p$, where $n_p$ is the concentration
of impurities. A linear suppression of $T_c$ is expected, and indeed
was observed experimentally. The parameter $\alpha_p$ computed from
the linear fit to the data, is plotted as a function of the film
resistance in Fig. 5.

\begin{figure}[h]
\begin{center}
\includegraphics[width = 3.0in]{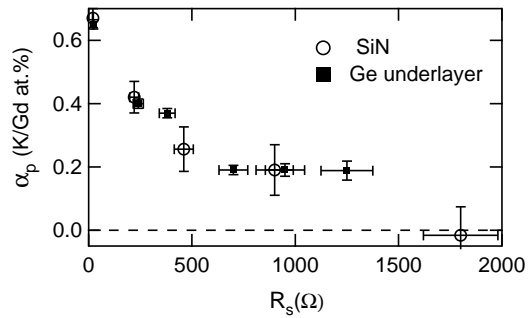}
\caption{\label{fig:TcGd2} The pair-breaking strength  per impurity,
$\alpha_p$, as a function of the film resistance for films deposited
on SiN and amorphous Ge}
\end{center}
\end{figure}

We found that with decreasing film thickness the pair-breaking
strength in MoGe films deposited on Ge drops by about three times
initially and saturates in films with thickness below 1.5 nm. From
the fermionic theories we expect that $\alpha_p\approx const$;
however this conclusion is made under assumption that the exchange
coupling between a localized spin and conduction electrons doesn't
change with decreasing film thickness or increasing disorder.

The behavior of $\alpha_p$ in MoGe films appears to be qualitatively
similar to the reduction of the Kondo contribution in thin films of
normal metals doped with magnetic atoms \cite{Giordano}. Extensive
studies of this effect revealed that it is stronger for impurities
with integer spin \cite{Giordano1} and depends on surface roughness
\cite{Giordano2}. The effect was explained in terms of spin-orbit
induced magnetic anisotropy for magnetic impurities in proximity to
the film surface \cite{Zawadowski}. The theory predicts that close
to the surface the effective spin of a magnetic impurity is reduced.

Both the Kondo effect and magnetic pair-breaking depend on total
impurity spin and exchange interaction between this spin and
conduction electrons. From the known diffusion coefficient $D=0.5$
cm$^2$/s \cite{Graybeal} and spin-orbit scattering time
 $\tau_{so}=5\times10^{-14}$ s \cite{Rogachev2} we can estimate
the average spin-orbit scattering length in MoGe as
$\ell_{so}=\sqrt{D \tau_{so}}\approx1.6$ nm. Experimentally, the
$\ell_{so}$ coincides with the film thickness below which the
saturation of $\alpha_p$ takes place. This observation suggests that
the pair-breaking strength of a Gd atom is reduced when it is
located within $\ell_{so}$ from the surface of a film; the growth of
$\alpha_p$ in thicker films corresponds to the increasing fraction
of Gd atoms with bulk-like surrounding. In other words, we have a
gradual transition form anisotropic to isotropic exchange.

For all MoGe films deposited on Ge we found that the suppression of
the superconductivity by the magnetic impurities and thickness
reduction are additive processes. A magnetic impurity introduced
into a superconductor suppresses the order parameter locally
\cite{Balatsky}. The local suppression of the order parameter with
decreasing film thickness is also a feature of the fermionic
mechanism of the SIT. In this regard, the additivity of the two
processes even very close to the critical point of the SIT is
consistent with the fermionic mechanism of the superconductivity
suppression. Moreover, the $\alpha_p \approx const$ relation that we
found in our thinnest films agrees with the specific prediction made
within the fermionic model.

Let us now discuss how the magnetic doping affects thin A-type
films. As shown in Fig. 4B, in the film with nominal thickness of 1
nm, the pair-breaking strength at low doping becomes zero; adding
magnetic impurities to the film does not change its $T_c$.  One
possibility for this anomalous response is that the spin-orbit
induced magnetic anisotropy gets stronger in films deposited on SiN,
because it has larger semiconductor gap. It is also possible that
the anomalous response to the magnetic doping is related to the
enhanced roughness of the thin A-type films, which may result in
inhomogeneous superconducting state. Spacial non-uniformity of the
order parameter is a common ingredient of the bosonic models.
Analysis of the effect of magnetic impurities within these models
can perhaps explain our findings.

In summary, we have studied the effect of magnetic doping on
superconducting Mo$_{70}$Ge$_{30}$ films. For uniform films
deposited on amorphous Ge, the suppression of superconductivity is
consistent with the fermionic mechanism. In thin films deposited on
SiN, the pair-breaking strength becomes zero. Further analysis is
needed to explain this anomalous response.

We thank L.N. Bulaevskii, A.M. Finkel'stein, P.M. Goldbart, L.
Ioffe, A. Kapitulnik, G. Refael,  and B. Spivak for discussions and
valuable comments. This work is supported by NSF CAREER Grant DMR
0955484.


\begin{thebibliography}{99}
\small

\bibitem{Poindexter} E. H. Poindexter,  \textit{Semiconductor Interfaces,
Microstructure,
and Devices}  (ed. Feng, Z. C.) Ch. 10, (Inst. Phys. Publ., Bristol,
Philadelphia, 1993).

\bibitem{Bluhm}
H. Bluhm, J.A. Bert, N.C. Koshnick, M.E. Huber, and K.A. Moler,
Phys. Rev. Lett. \textbf{103}, 026805 (2009).

\bibitem{McDermott}
S. Sendelbach, D. Hover, A. Kittel, M. M\"{u}ck, J.M. Martinis, and
R. McDermott, Phys. Rev. Lett. \textbf{100}, 227006 (2008). L. Faro
and L.B. Ioffe, Phys. Rev. Lett. \textbf{100} 227005 (2008); S.
Choi, D.-H. Lee, S. G. Louie, and J. Clarke, Phys. Rev. Lett.
\textbf{103}, 197001 (2009); F. Yoshihara, Y. Nakamura, and J.S.
Tsai, Phys. Rev. B \textbf{81}, 132502 (2010).

\bibitem{Rogachev}
A. Rogachev, T.-C. Wei, D. Pekker, A.T. Bollinger, P.M. Goldbart,
and A. Bezryadin, Phys. Rev. Lett. \textbf{97}, 137001 (2006).

\bibitem{Heviland}
D.B. Haviland, Y. Liu, and A.M. Goldman, Phys. Rev. Lett.
\textbf{62}, 2180 (1989);

\bibitem{Markovic1}
N. Markovic, C. Christiansen, and A.M. Goldman, Phys. Rev. Lett.
\textbf{81}, 5217 (1998).

\bibitem{Gantmakher}
V.F. Gantmakher, M.V. Golubkov, V.T. Dolgopolov, G.E. Tsydynzhapov,
and A.A. Shashkin, JETP Lett. \textbf{68}, 363 (1998);

\bibitem{Shahar}
G. Sambandamurthy, L.W. Engel, A. Johansson, and D. Shahar, Phys.
Rev. Lett. \textbf{92}, 107005 (2004).

\bibitem{Baturina1}
T.I. Baturina, A.Yu. Mironov, V.M. Vinokur, M.R. Baklanov, and C.
Strunk, Phys. Rev. Lett. \textbf{99}, 257003 (2007).

\bibitem{Kapitulnik1}
M.A. Steiner, N.P. Breznay, and A. Kapitulnik, Phys. Rev. B
\textbf{77}, 212501 (2008).

\bibitem{Finkelstein1}
A.M. Finkel'stein, JETP Lett. \textbf{45}, 46 (1987).

\bibitem{Kirkpartick}
T.R. Kirkpatrick and D. Belitz, Phys. Rev. Lett. \textbf{68}, 3232
(1992).

\bibitem{Belitz}
T.P. Devereaux and D. Belitz, Phys. Rev. B \textbf{35}, 359 (1996);

\bibitem{Smith}
 R.A. Smith and V. Ambegaokar, Phys. Rev. B \textbf{62}, 5913
(2000).

\bibitem{Parker1}
J.S. Parker, D.E. Read, A. Kumar, and P. Xiong, Europhys. Lett.
\textbf{75}, 950 (2006).

\bibitem{Chervenak1}
J.A. Chervenak and J.M. Valles, Jr., Phys. Rev \textbf{51}, 11977
(1995).

\bibitem{Fisher1}
M.P.A. Fisher, Phys. Rev. Lett. \textbf{65}, 923 (1990).

\bibitem{Feigelman1}
M.V. Feigel'man, L.B. Ioffe, V.E. Kravtsov, and E.A. Yuzbashyan,
Phys. Rev. Lett. \textbf{98}, 027001 (2007).

\bibitem{Pokrovsky1}
V.L. Pokrovsky, G.M. Falco, and T. Nattermann, Phys. Rev. Lett.
\textbf{105}, 267001 (2010).

\bibitem{Spivak1}
B.I. Spivak and S.A. Kivelson, Phys. Rev. B \textbf{43}, 3740
(1991).

\bibitem{Ghosal}
A. Ghosal, M. Randeria, and N. Trivedi, Phys. Rev. B \textbf{65},
014501 (2001).

\bibitem{Sacepe1}
B. Sacepe \textit{et al.}, Phys. Rev. Lett. \textbf{101}, 157006
(2008); B. Sacepe \textit{et al.}, Nat. Phys. \textbf{7}, 239
(2011).

\bibitem{Graybeal}
J.M. Greybeal, PhD thesis, Stanford University (1985).

\bibitem{Altshuler}
B.L. Altshuler and A.G. Aronov, in \textit{Electron-Electron
Interaction in Disordered Systems}, eds. A.L. Efros and M. Pollak,
Elseveir Science Publ. (1985).

\bibitem{Herzog}
A.V. Herzog, P. Xiong, F. Sharifi, and R.C. Dynes, Phys. Rev. Lett.
\textbf{76}, 671 (1996).

\bibitem{AG}
A. A. Abrikosov and L. P. Gor'kov, Sov. Phys. JETP \textbf{15}, 752
(1962).

\bibitem{Giordano}
G. Chen and N. Giordano, Phys. Rev. Lett. \textbf{66}, 209 (1991);
J.F. DiTusa, K. Lin, M. Park, M.S. Isaacson, and J.M. Parpia, Phys.
Rev. Lett. \textbf{68}, 1156 (1992).

\bibitem{Giordano1}
T.M. Jacobs and N. Giordano, Europhys. Lett. \textbf{44}(1) 74
(1998).

\bibitem{Giordano2}
V. Chandrasekhar \textit{et al.}, Phys. Rev. Lett. \textbf{72}, 2053
(1994).  N. Giordano and T.M. Jacobs, J. Appl. Phys. \textbf{87},
6079 (2000).

\bibitem{Zawadowski}
O.\'{U}js\'{a}ghy, A. Zawadowski, and B.L. Gyorffy, Phys. Rev. Lett.
\textbf{76}, 2378 (1996).

\bibitem{Rogachev2}
A. Rogachev, A.T. Bollinger, and A. Bezryadin, Phys. Rev. Lett.
\textbf{94}, 017004 (2005).

\bibitem{Balatsky}
A.V. Balatsky, I. Vekhter, and J.-X. Zhu, Rev. Mod. Phys.
\textbf{78}, 373 (2006).

\end{thebibliography}
\end{document}